\documentclass[11pt,a4paper]{article}

\usepackage[utf8]{inputenc}
\usepackage[T1]{fontenc}
\usepackage{amsmath,amssymb}
\usepackage{graphicx}
\usepackage{hyperref}
\usepackage{geometry}
\title{\bf{Ugo Amaldi, the DELPHI Collaboration and}\\\bf{the Physics Legacy of LEP}}

\author{Alessandro De Angelis\\[4pt]
\small Physics and Astronomy Department ``Galileo Galilei'' of the University of Padova,\\
\small via Marzolo 8, 35131 Padova (Italy)\\
\small and\\
\small Permanent Delegation of Italy to the International Organizations,\\
\small rue de Varenne 50, 75007 Paris, France\\[4pt]
\small E-mail: \texttt{alessandro.deangelis@unipd.it}\\
\small ORCID: 0000-0002-3288-2517}

\date{}

\begin{document}

\maketitle

\begin{abstract}
\noindent This article offers a portrait of the DELPHI experiment at CERN's Large Electron--Positron Collider (LEP) through the scientific life and leadership of Ugo Amaldi. It traces how DELPHI contributed to LEP's physics program, from precision studies at the Z pole to higher-energy running with W-pair production and increasingly ambitious Higgs searches. Along the way, it highlights Ugo Amaldi's technical and organizational innovations, especially his insistence on bold detector choices and his sustained support for young physicists and collaborative leadership. The article also recalls Ugo's influential work on the unification of forces and shows how DELPHI's technologies, software, governance structures, and data-sharing practices anticipated many features of later collider experiments and of contemporary science policy. In this sense, DELPHI's legacy is not only foundational for today's particle physics, but also a lasting and formative element of modern scientific culture.
\end{abstract}

\noindent\textbf{Keywords:} CERN, Ugo Amaldi, Gauge Bosons, Higgs boson searches, Precision electroweak measurements, Artificial Intelligence in Physics, Open Science
%
%\selectlanguage{italian}
%
%\section*{Riassunto}
%Questo articolo parla dell'esperimento DELPHI al Large Electron--Positron Collider (LEP) del CERN attraverso la vita scientifica e la leadership di Ugo Amaldi. Ripercorre i risultati di DELPHI dagli studi di precisione al picco della Z fino ai dati a energie pi? elevate, con la produzione di coppie di bosoni W e la ricerca di nuove particelle, in particolare del bosone di Higgs. Lungo questo percorso mette in evidenza le innovazioni tecniche e organizzative introdotte da Ugo Amaldi, in particolare le sue coraggiose scelte progettuali per il rivelatore, il costante sostegno ai giovani e la promozione di uno stile collaborativo di gestione. L'articolo richiama inoltre il ruolo di Ugo nello studio dell'unificazione delle forze e mostra come le tecnologie, il software, le strutture di governance e le pratiche di condivisione dei dati di DELPHI abbiano anticipato molte caratteristiche dei successivi esperimenti contribuendo a plasmare la moderna politica della scienza. In questo senso, l'eredit? di DELPHI non ? soltanto fondamentale per la fisica delle particelle di oggi, ma costituisce anche un elemento duraturo e formativo della cultura scientifica moderna.
%
%\noindent\textbf{Parole chiave:} CERN, Ugo Amaldi, Bosoni di gauge, Ricerca del bosone di Higgs, Misure elettrodeboli di precisione, Intelligenza artificiale in fisica, Open Science

%\selectlanguage{english}

\section{Introduction}

The Large Electron--Positron Collider (LEP) at CERN, which operated between 1989 and 2000 \cite{ref1,ref2}, was conceived with three main goals, in order of ambition: to reveal an abundance of new, previously unknown particles; to discover the Higgs boson; and to probe with unprecedented precision the structure of the Standard Model, in particular the properties of the Z and W bosons. The second goal was narrowly missed -- although LEP's indirect determination of the Higgs boson mass, obtained years before the start of the LHC, was fully consistent with the first direct measurement. About the third goal, LEP's achievements ushered in a new era of precision particle physics, and its legacy continues to define the field \cite{ref3,ref4}.

Among the four major LEP experiments -- ALEPH, DELPHI, L3 and OPAL -- DELPHI was in many respects the most ambitious and intricate. At the heart of DELPHI's story stands Ugo Amaldi, a scientist whose career bridges high-energy physics, the physics of accelerators and their medical applications, scientific politics and physics teaching. Trained in Rome, Ugo worked at CERN for decades and became the driving force behind DELPHI, serving as its spokesperson for thirteen years. In this article, through the eyes of Ugo's friends and collaborators in DELPHI, I retrace the experiment's technical achievements, the style of scientific leadership that shaped the collaboration, and the enduring legacy of DELPHI -- inseparable from that of LEP as a whole.

\section{From COLLEPS to DELPHI: The Game of a Name}

Ugo Amaldi started collecting a group of scientists for a new experiment at the LEP collider, then in construction, in 1980. In the beginning, the collaboration that would eventually become DELPHI was called COLLEPS: COLlaboration for LEP Studies. It was soon agreed that this label was, to put it gently, lacking in inspiration. A new acronym was sought, and ``DELPHI'' was finally adopted: DEtector with Lepton, Photon and Hadron Identification. The name was proposed by Gerald Myatt, long-serving Chairman of the Collaboration Board of the experiment, which was approved by the LEP Experiments committee after a presentation by Ugo to the LEP Experiment Committee in 1982 (Fig.~1). At the end of that year, I started working part-time in DELPHI as an undergraduate.

\begin{figure}

\centering\includegraphics[width=0.9\linewidth]{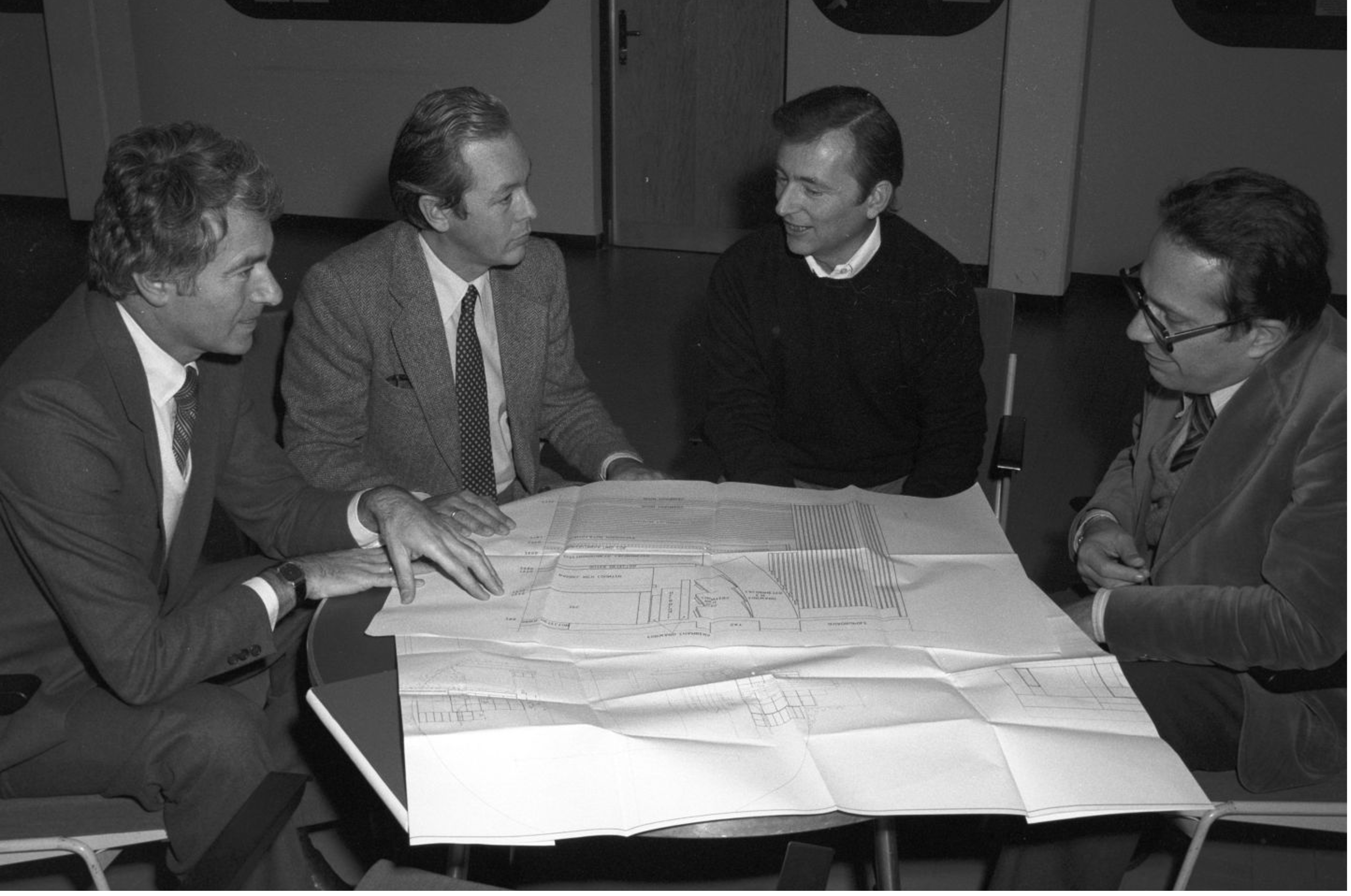}

\caption{Discussing the design of DELPHI in 1982. From left to right: Gregoire Kantardjian, Ugo Amaldi, Hans-Jurgen Hilke (DELPHI technical coordinator), Guido Petrucci. Courtesy of CERN.}

\end{figure}

With the new name came the inevitable question of a logo. The selected design, proposed by Angelo Rampazzo from Padua, drew on the mythological link between Apollo and Delphi, the town where an Oracle revealed to humankind previously unknown truths. In one version of the Greek myth, Apollo travels to Delphi in the form of a dolphin (\emph{delfino} in Italian), carrying gifts. To make the reference even clearer, the DELPHI dolphin carried a stylized ``Z'' on its body (Fig.~2).

\begin{figure}[h]

\centering\includegraphics[width=0.2\linewidth]{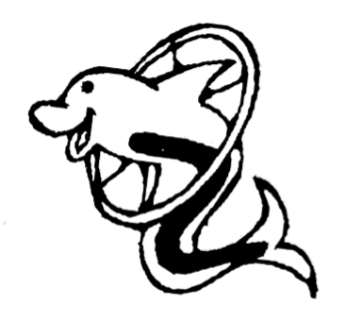}

\caption{The DELPHI logo.}

\end{figure}

More important than the logo was a fundamental strategic decision. As early discussions made clear, the nascent collaboration faced a classic dilemma: should one build a conservative detector, safe and robust, or instead venture towards a more daring design, opening new horizons in particle identification and vertex reconstruction? Ugo Amaldi firmly championed the latter course. Under his influence, DELPHI committed itself to highly granular tracking, sophisticated particle-identification systems and a powerful vertex detector -- an ambitious combination that would later prove decisive for b- and c-quark physics.

Experts in the then-innovative Ring Imaging Cherenkov (RICH) technique -- Tom Ypsilantis, Jacques S\'eguinot and, later, Tord Ekel\"of -- joined the collaboration, integrating advanced Cherenkov-based particle identification at the heart of the detector concept. At the same time, Ugo insisted that the innermost region preserve enough space for three layers of silicon strip detectors. This choice embodies a simple geometrical truth: to say it with a joke, only one straight line passes through three experimental points. With three silicon layers, DELPHI could reconstruct charged-particle tracks from the primary interaction point and from secondary decay vertices with excellent precision -- a crucial asset for tagging b-quark decays at a time when the b-quark lifetime was still poorly known.

The main tracking device selected was a large time projection chamber (TPC), surrounded by the RICH, all enclosed within what was then the largest superconducting coil ever built: approximately 7.4\,m in length, 6.2\,m in diameter, and providing a magnetic field of 1.2\,tesla. Combined with additional tracking systems and the calorimeters, this configuration made DELPHI (Fig.~3) not merely a tracking detector, but a true three-dimensional camera for particle collisions, endowed with powerful capabilities for particle identification \cite{ref5}.

\begin{figure}

\centering\includegraphics[width=0.9\linewidth]{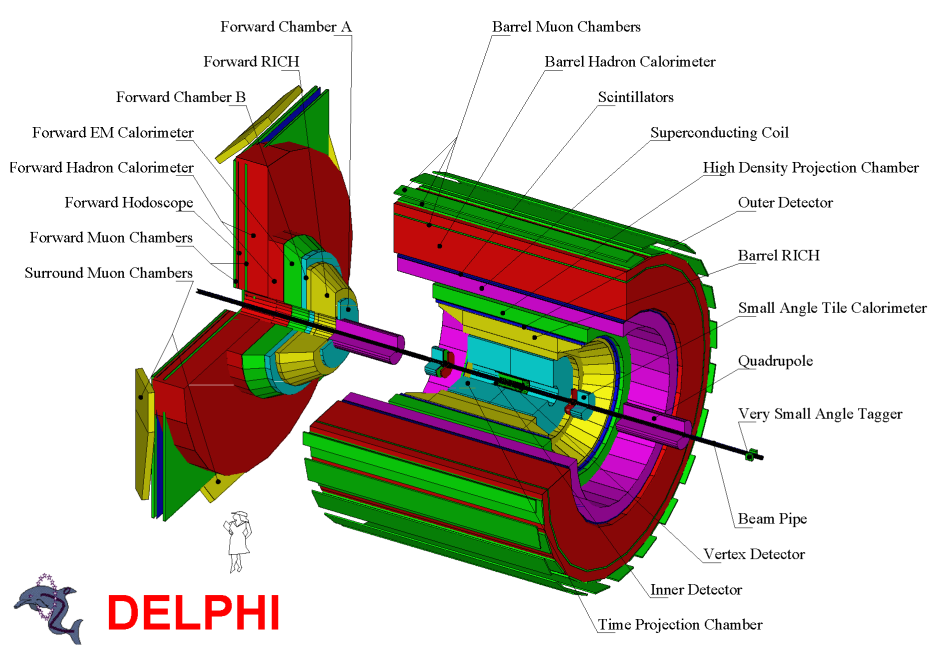}

\caption{The DELPHI detector. It looks so nice that it has been used by Macao for a stamp!}

\end{figure}

\section{Building and Commissioning: Stress Handling, Yellow Tape, and the First Z}

Construction of DELPHI began in the early 1980s and unfolded over several years, in parallel with the construction of the LEP ring itself. The installation schedule was tight and often stressful; I remember Ugo Amaldi sleeping only a few hours a day at the Novotel near CERN and spending the rest of his time in the control room down in the cavern, whose organized chaos was reminiscent of Dante's description of the Venetian Arsenal in the \emph{Inferno}. At one point, Ugo laid a strip of yellow tape across the control-room floor, and only those whom he explicitly invited to cross it were allowed into the central area. It was a visible reminder both of the discipline required during critical operations and of the strong, personal style of leadership that held this extraordinarily complex project together.

In July 1989 LEP finally began operation. DELPHI was ``more or less'' ready when the beams were first circulated, and in August the first events began to appear on the displays. But on 14 August, while the other three LEP experiments had already recorded clear Z-boson events, DELPHI still had none. This absence led to some tense hours: after all the effort, could something be fundamentally wrong with the detector? In the end, the problem turned out not to lie with us but with the machine: a misalignment of the beams in the DELPHI interaction region. Once corrected, DELPHI too started to see beautiful Z peaks, and before the end of the year it was fully integrated into the common physics program.

During its first months of running (Fig.~3), LEP collected data at five different center-of-mass energies, spaced by about 1\,GeV around the Z resonance. This ``energy scan'' allowed a precise determination of the Z mass and width and, crucially, of the number of light neutrino species. By comparing the total invisible decay width of the Z -- essentially, how often it decays into particles that leave no direct trace in the detectors -- with the width expected from a single neutrino type, the LEP experiments showed that there are three, and only three, families of light neutrinos in nature (Fig.~4). This result, obtained after only a few days of operation, remains one of the enduring legacies of LEP.

\begin{figure}

\centering\includegraphics[width=0.5\linewidth]{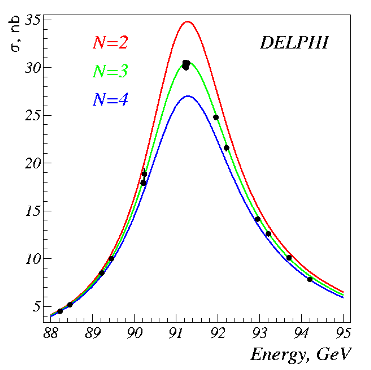}

\caption{Z cross section versus center-of-mass energy as measured by DELPHI, compared to what expected for 2, 3, and 4 light neutrino species.}

\end{figure}

The first DELPHI physics paper, on the measurement of the Z mass and width from multi-hadronic final states \cite{ref6}, has an amusing backstory. Eager to be ready for rapid publication, Ugo urged that the analysis and a draft paper be prepared even before the first collisions were observed -- a course of action that, for Italian physicists, carried an unmistakable whiff of bad luck. The draft, written before knowing anything about the real performance of the detector, contained the cheerful sentence: ``the trigger efficiency is believed to be $\left(98 \pm 2\right)\%$ \ldots''. This bold estimate survived all subsequent revisions and, in that exact form, made it into the published, refereed, version.

\section{Handling Complexity: Necessity is the Mother of Invention}

As already noted, DELPHI was famously complex. Nearly twenty subdetectors were integrated into a central cylindrical structure with endcaps, providing detailed measurements over almost the full solid angle. The geometry was intricate, and the information it produced was both widely distributed and highly heterogeneous: tracking detectors, calorimeters, RICH counters, muon chambers, time-of-flight systems, and more.

This complexity forced the collaboration to be ambitious on the software side. Enormous effort went into developing simulation, reconstruction, and analysis tools capable of fusing the many different detector responses into coherent physics observables. This included graphical interfaces for event and detector-performance visualization (Fig.~5), structured databases for conditions, calibrations and configurations, flexible analysis frameworks, and -- above all -- sophisticated particle-identification algorithms that combined inputs from the trackers, RICH, calorimeters and muon systems. Much of the credit for these achievements goes to the long-time software coordinator, Luc Pape.

\begin{figure}

\centering\includegraphics[width=0.6\linewidth]{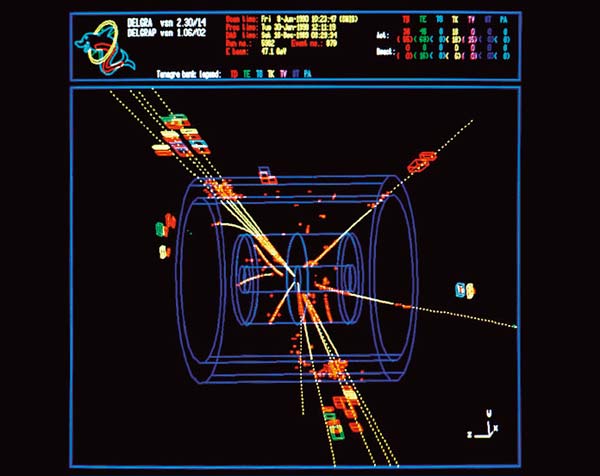}

\caption{A Z decay into two jets and a muon-antimuon pair.}

\end{figure}

From the quality of the detector and from this work emerged one of DELPHI's most distinctive achievements: highly efficient and remarkably pure tagging of b- and c-quark jets. A flagship analysis, classifying hadronic Z decays into b- and c-quark pairs, used artificial neural networks -- the first published high-energy physics analysis \cite{ref7} based on artificial intelligence, now a standard tool in particle physics (and well beyond). In the early 1990s this was a bold and pioneering choice, accompanied by intense discussions on issues such as reliability, robustness, and reproducibility. Ugo Amaldi was always open to the new and encouraging courageous attitudes.

DELPHI's excellent heavy-flavor tagging capability underpinned not only precision measurements of the Z couplings to b and c quarks, but also the combined LEP/SLD electroweak fits that indirectly constrained the masses of the top quark and, later, the Higgs boson. Many of these results, even decades on, remain benchmark measurements in particle physics.

\section{Leadership and Collaboration: Giving Visibility}

By 1993, Ugo Amaldi had served as spokesperson of DELPHI for thirteen years, the last four devoted to data taking. Aware both of the institutional need for turnover and of the importance of creating space for colleagues who had earned it, he urged the collaboration to adopt a new governance scheme. From that point on, DELPHI would be led by a Spokesperson with a two-year mandate, alongside a Spokesperson-Elect who would subsequently assume the role. Over the following years, the role of Spokesperson was held in turn (Fig.~6) by Jean-Eudes Augustin, Daniel Treille, Wilbur Venus, Tiziano Camporesi and Jan Timmermans, but the structures introduced under Ugo's leadership remained in place.

\begin{figure}

\centering\includegraphics[width=0.5\linewidth]{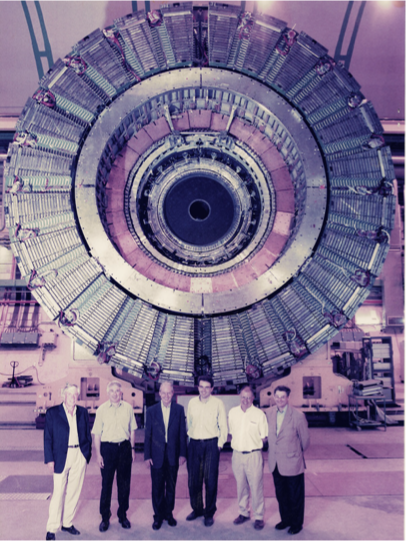}

\caption{From left to right: Daniel Treille, Jan Timmermans, Ugo Amaldi, Tiziano Camporesi, Wilbur Venus and Jean-Eudes Augustin.}

\end{figure}

At Ugo's instigation, DELPHI also developed mechanisms to give younger collaborators greater visibility, notably by promoting the nomination of early-career physicists to present key results and by allowing the publication of internal notes in which the real driving forces behind each result could be clearly identified. Ugo has always been extremely generous in supporting his collaborators even after their time in DELPHI, as I can personally testify, not only by recommending them, but, even more importantly, by offering wise advice: in a word, by teaching.

An important organizational innovation was the strong support for LEP-wide working groups. These groups combined results from the four experiments to produce joint measurements with properly treated correlated systematics -- a non-trivial statistical and sociological challenge. Ugo and others argued that this work should be carried out by experimentalists directly involved in the detectors and analyses, rather than being delegated solely to external committees. The LEP Electroweak Working Group and similar efforts went on to become templates for later combinations at the Tevatron and the LHC.

DELPHI also argued, relatively early on, for structured access to its data by scientists outside the collaboration, under clearly defined rules. At the time this was a forward-looking position; only in recent years, with the rise of data-preservation initiatives in high-energy physics, has such external access become standard practice. In this respect, DELPHI anticipated the modern movement towards Open Science.

\section{Energy Upgrades: From the Z Pole to WW Pairs and Higgs Hunting}

After several years of running near the Z resonance, LEP progressively increased its center-of-mass energy. In July 1996, the collider reached 161\,GeV, just above the threshold for producing pairs of W bosons. On the first day of regular run, DELPHI recorded the first WW event at LEP, visible in the detector through one of the characteristic (4-jet) topologies (Fig.~7). This event was reported in a feature article by the \emph{Science} magazine.

\begin{figure}

\centering\includegraphics[width=0.5\linewidth]{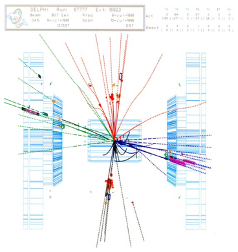}

\caption{From DELPHI Event visualization. Reproduced in A.~Watson, Science 273 (1996) 306.}

\end{figure}

The higher-energy LEP (frequently called LEP2) program opened the way to precision measurements of the W mass and width, detailed tests of the triple gauge-boson couplings, and searches for new particles, including the Standard Model Higgs boson. DELPHI played a full part in this program, exploiting its excellent tracking and particle-identification capabilities to unravel complex final states.

By the year 2000, LEP had been pushed to center-of-mass energies of up to about 209\,GeV, close to the maximum allowed by the accelerator design and by the length of the LEP tunnel. In that final year of operation, analyses in some LEP experiments began to show tantalizing hints of an excess of Higgs candidates around a mass of 114\,GeV, near the upper edge of the kinematically accessible range. The statistical significance, however, was not decisive, and opinions in the community were divided. Some physicists (including the author of this article) simply did not believe that the evidence was real; others judged it insufficient to justify delaying the planned transition to the Large Hadron Collider (LHC); still others argued that one or two additional years of running at slightly higher energy could either confirm or refute the signal.

The decision rested with CERN's Director-General at the time, Luciano Maiani. Extending LEP would have meant postponing the dismantling of the machine and the start of LHC construction in the same tunnel, incurring substantial financial penalties for breaking industrial contracts and risking a loss of momentum for the LHC project and its possible cancellation. In the end, Maiani decided not to prolong LEP, and the collider was shut down in November 2000 \cite{ref1,ref2,ref8}.

In hindsight, we know that the Higgs boson has a mass of about 125\,GeV, as discovered by ATLAS and CMS in 2012 at the LHC. That value lay beyond the direct reach of LEP at the maximum energy it ultimately achieved; however, with new radiofrequency cavities, the LEP center-of-mass energy could have been pushed to around 220\,GeV \cite{ref2}, as asked by some, including -- insistently -- Daniel Treille, also before the final phase of the accelerator. A 125\,GeV Higgs might have been observed at LEP after a long run at such energies. That said, the demands of such an upgrade of the machine, and the corresponding extension of the experiments, would in many respects have been extremely high, possibly too high.

\section{Unification and Beyond: An Image Worth a Thousand Words}

One of the most widely cited papers associated with Ugo Amaldi's name does not stem from a specific DELPHI analysis, but from his broader interest in the unification of forces. In a 1991 paper with Wim de Boer and Hermann F\"urstenau, based on an idea of his, the then-new precision measurements of the electroweak and strong coupling constants -- many of them obtained at LEP and by DELPHI in particular -- were used to study how these couplings evolve with energy under the renormalization group \cite{ref9}.

By plotting the inverses of the three Standard Model gauge couplings (electromagnetic, weak and strong) as straight lines on a suitable logarithmic energy scale, Ugo and his collaborators showed that, within the minimal Standard Model, these three lines do not quite meet at a single point when extrapolated to very high energies. However, if one extends the model to include supersymmetric partners of the known particles in a mass range around 1\,TeV (a value enjoying the qualities of being both appropriate and testable), the three lines converge strikingly well at a common unification scale around $10^{15}$--$10^{16}\,\mathrm{GeV}$ (Fig.~8).

\begin{figure}

\centering\includegraphics[width=\linewidth]{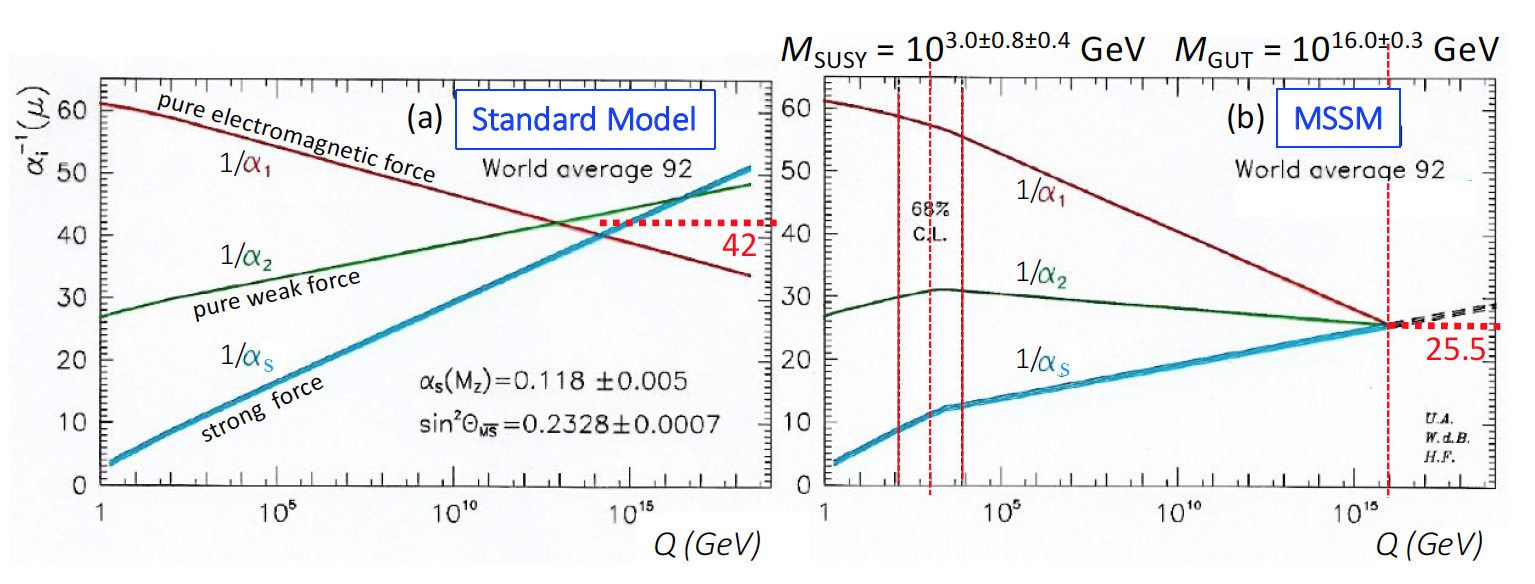}

\caption{Unification plots computed with the best experimental values in summer 1992 \cite{ref9,ref10}.}

\end{figure}

The visual power of that plot -- three almost straight lines intersecting in a single point in the supersymmetric case -- together with the novelty of quoting fitted values for both the unification scale and a characteristic supersymmetric mass scale, made a deep impression. It quickly became, and still remains, a favorite figure in particle-physics lectures. Although experimental searches have not (yet) revealed supersymmetric particles, the idea that gauge couplings might unify remains central to many theories beyond the Standard Model; Ugo Amaldi's plot is still frequently shown in discussions of grand unification at every level, from specialist seminars to popular talks, as a way to illustrate supersymmetry as a possible key to what lies behind the door.

Beyond unification, DELPHI and LEP more broadly helped to shape the landscape of ``New Physics'' precisely through what they did not see. Precision electroweak fits, incorporating millions of Z decays and hundreds of thousands of W events, placed stringent constraints on possible deviations from the Standard Model, indirectly limiting the masses and couplings of hypothetical heavy particles and interactions. In this sense, LEP and DELPHI played a role analogous to that of an ultra-fine microscope: they did not uncover a zoo of entirely new objects, but they revealed with exquisite clarity the structure and internal consistency of the theory we already possessed, locating many conceivable new phenomena -- including popular classes of Dark-Matter candidates, central to today's astrophysical research -- to well defined energy scales.

\section{The Legacy of DELPHI and LEP}

By the time LEP was dismantled and its tunnel repurposed for the LHC, DELPHI had accumulated a vast dataset and a broad portfolio of results: precision measurements of Z and W properties, studies of heavy flavors, searches for new particles, tests of QCD in $e^{+}e^{-}$ annihilation, and much more \cite{ref10}. Its technological contributions -- especially in silicon vertexing, RICH particle identification, large-scale superconducting magnets and sophisticated offline software -- influenced the designs of later experiments at the Tevatron and LHC.

Although, as already mentioned, DELPHI excelled in particle identification -- which made its results particularly prominent in QCD and in b physics -- the scientific legacy of each individual LEP detector is, 25 years on, difficult to disentangle, and LEP must be regarded as a whole. LEP's scientific legacy is to have transformed the Standard Model into a quantitatively tested ``working theory of matter'' by delivering per-mille-level measurements of key electroweak observables (gauge-boson masses and widths, couplings, the number of light neutrino species and heavy-flavor parameters), precise determinations of the strong coupling and its energy evolution, and clear evidence for the running of masses and couplings \cite{ref3,ref4,ref10}.

Organizationally, DELPHI left a legacy in how large collaborations manage authorship, conference talks, data combinations and, more recently, data preservation. Many of the practices that are now standard in big experiments were tested and refined in the LEP era.

For Ugo Amaldi, DELPHI was the longest chapter on fundamental physics in a broader career that later turned strongly toward medical physics, particularly hadron therapy for cancer treatment. Yet, as his 90th-birthday symposium makes clear, his years as DELPHI spokesperson and architect of its scientific strategy remain a central part of his contribution to science. The combination of technical insight, willingness to take calculated risks, attention to young collaborators and openness to innovation (including early adoption of artificial intelligence and advocacy for data sharing) set a tone.

Today, when we look at the discovery of the Higgs at the LHC or at ongoing efforts to clarify the physics beyond the Standard Model, it is easy to focus on the newest machines and analyses. But the intellectual and technical foundations of much of that work were laid at LEP, and DELPHI was probably the most creative and influential among the experiments. Scientific progress is not only about spectacular discoveries, but also about careful craftsmanship, bold design choices and the patient accumulation of precise knowledge.

In that sense, ``long live DELPHI'' as a living part of the culture and history of high-energy physics, and, of course, long live Ugo!

\newpage

\section*{Acknowledgements}

I acknowledge Gerald Myatt for providing me material and checking the final text of this article. I thank Ugo for everything. This article is dedicated to the members of the DELPHI Collaboration who are no longer with us.

\end{document}